\documentclass[a4paper,fleqn,usenatbib]{mnras}

\usepackage{footnote}
\usepackage[T1]{fontenc}
\usepackage{ae,aecompl}

\usepackage{graphicx}	
\usepackage{amsmath}	
\usepackage{amssymb}	
\newcommand{\kms}{km\,s$^{-1}$} 

\newcommand{\hi}{H{\sc i}}
\newcommand{\hii}{H{\sc i}~21\,cm}

\title[Associated H{\sc i} 21\,cm absorption at $z = 1.223$]{Giant Metrewave Radio Telescope detection of associated 
H{\sc i} 21\,cm absorption at $z = 1.2230$ towards TXS\,1954+513}

\author[Aditya et al.]{J. N. H. S. Aditya$^{1}$\thanks{adityaj@iucaa.in},
Nissim~Kanekar$^1$\thanks{Swarnajayanti Fellow}, J. Xavier Prochaska$^2$, Brandon Day$^3$,
\newauthor 
Paul Lynam$^2$, Jocelyn Cruz$^3$\\
$^{1}$National Centre for Radio Astrophysics, Tata Institute of Fundamental Research, Pune 411007, India\\
$^{2}$UCO/Lick Observatory, University of California -- Santa Cruz, Santa Cruz, CA 95064, USA \\
$^{3}$University of California -- Santa Cruz, Santa Cruz, CA 95064, USA}

\date{Accepted XXX. Received YYY; in original form ZZZ}

\pubyear{2016}

\begin{document}
\label{firstpage}
\pagerange{\pageref{firstpage}--\pageref{lastpage}}
\maketitle

\begin{abstract}
We have used the 610~MHz receivers of the Giant Metrewave Radio Telescope (GMRT) to detect associated 
H{\sc i} 21\,cm absorption from the $z = 1.2230$ blazar TXS\,1954+513. The GMRT H{\sc i} 21\,cm absorption 
is likely to arise against either the milli-arcsecond-scale core or the one-sided milli-arcsecond-scale 
radio jet, and is blueshifted by $\approx 328$~km~s$^{-1}$ from the blazar redshift. This is consistent with a 
scenario in which the H{\sc i} cloud giving rise to the absorption is being driven outward by the radio jet. The 
integrated H{\sc i} 21\,cm optical depth is $(0.716 \pm 0.037)$~km~s$^{-1}$, implying a high H{\sc i} column density, 
$N_{\rm HI} = (1.305 \pm 0.067) \times ({\rm T_s/100\: K}) \times 10^{20}$~cm$^{-2}$, for an assumed H{\sc i}
spin temperature of 100~K. We use Nickel Telescope photometry of TXS\,1954+513 to infer a high rest-frame 
1216~\AA\ luminosity of $(4.1 \pm 1.2) \times 10^{23}$~W~Hz$^{-1}$. The $z = 1.2230$ absorber towards 
TXS\,1954+513 is only the fifth case of a detection of associated H{\sc i} 21\,cm absorption at $z > 1$, 
and is also the first case of such a detection towards an active galactic nucleus (AGN) with a rest-frame 
ultraviolet luminosity $\gg 10^{23}$~W~Hz$^{-1}$, demonstrating that neutral hydrogen can survive in AGN 
environments in the presence of high ultraviolet luminosities.
\end{abstract}



\section{INTRODUCTION}

``Associated'' \hii\ absorption studies have been used to probe physical conditions 
in neutral atomic hydrogen (\hi) in the environments of active galactic nuclei (AGNs) 
for more than four decades \citep[e.g.][]{roberts70}. 
Such studies allow one to trace the presence and kinematics of, as well as physical conditions 
in, \hi\ in AGN environments as a function of redshift and AGN type. For example, comparisons 
between the \hii\ absorption redshift with the AGN emission redshift allow one to determine 
whether the detected absorption arises from gas outflow or infall, and thus to study the 
fuelling of the active nucleus or the effects of AGN feedback 
\citep[e.g.][]{morganti05,morganti13}.One can also use such observations, especially with 
Very Long Baseline Interferometric (VLBI) techniques, to study the kinematics of circumnuclear disks 
around AGNs \citep[e.g.][]{conway95,peck02}, interactions between the active 
nucleus, the radio jet and the ambient gas \citep[e.g.][]{morganti05b}, and even to infer 
the presence of binary super-massive black holes in active galaxies \citep[e.g.][]{morganti09}. 
The detectability of 
such associated \hii\ absorption in different AGN types can also be used to test AGN unification 
schemes \citep[e.g.][]{barthel89}. 

More than a hundred radio-loud AGNs have been searched for associated \hii\ absorption, 
from local systems out to very high redshifts, $z \approx 5.2$ 
\citep[e.g.][]{vangorkom89,vermeulen03,gupta06,carilli07,curran08,gereb15,aditya16}. 
More than 50 associated systems have so far been detected in \hii\ absorption till now, 
with the vast majority at low redshifts, $z \lesssim 0.7$ 
\citep[e.g.][]{vermeulen03,gupta06,gereb15}, and, indeed, only four confirmed detections 
at $z > 1$ \citep{uson91,moore99,ishwar03,curran13}. While this might suggest redshift 
evolution in the neutral gas content in AGN environments, the heterogeneity of most target 
samples, with typically a mix of compact and extended background radio sources, has meant 
that it has been difficult to separate the AGN characteristics from the above redshift 
evolution. It has also been pointed out that \hii\ absorption has never been detected 
in the environment of an AGN with a rest-frame ultraviolet (UV) luminosity $\gtrsim 10^{23}$~W~Hz$^{-1}$, 
at either low or high redshifts \citep{curran08}. These authors argue that the above 
luminosity may be a threshold above which neutral gas does not survive in AGN environments.

The two main problems with using \hii\ absorption to probe the redshift evolution of 
AGN environments have been the lack of searches for \hii\ absorption at high redshifts, $z > 1$,
and the heterogeneity in the target samples at all redshifts. We have hence been 
using the Giant Metrewave Radio Telescope (GMRT) to carry out a search for redshifted \hii\
absorption in a large, uniformly-selected sample of compact sources, the Caltech-Jodrell Flat-spectrum 
(CJF) sample \citep{pearson88,taylor96}. Our pilot study of a subset of this sample yielded 
the first statistically significant evidence for redshift evolution in the detectability of 
associated \hii\ absorption in a uniformly-selected sample \citep{aditya16}. However, the data 
are also consistent with a scenario in which AGNs with a high UV or radio luminosity have 
a lower \hii\ absorption strength, as suggested by \citet{curran08}. In the present 
{\it Letter}, we report a new detection of redshifted \hii\ absorption from our GMRT 
survey of flat-spectrum radio sources, at $z \approx 1.2230$ towards TXS\,1954+513, the 
first case of associated \hii\ absorption in a UV-luminous quasar.

%
%
%


\begin{figure}
\includegraphics[scale=0.45]{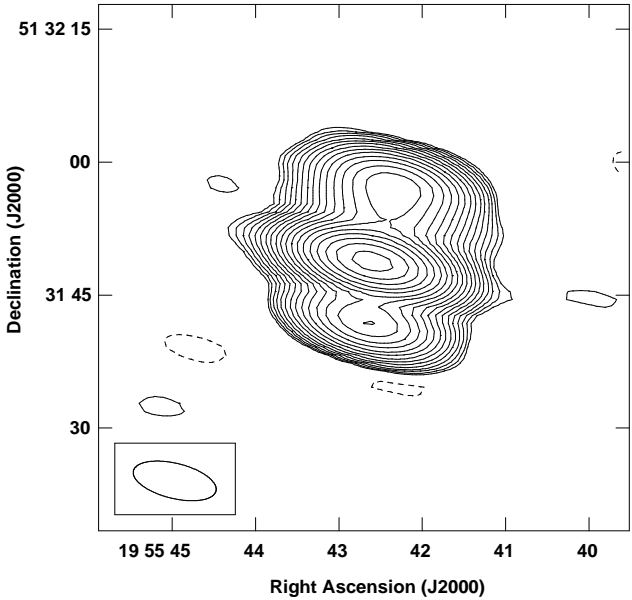}
\caption{The GMRT 640~MHz continuum image of TXS\,1954+513, showing the core-dominated triple 
structure. The positive contour levels extend from $2$~mJy to $512$~mJy, in steps of $\sqrt{2}$,
while the sole negative (dashed) contour is at $-2$~mJy. The peak continuum flux is 
$1.22130 \pm 0.00069$~Jy/Beam, and the synthesized beam has a full-width-at-half-maximum 
of $10.6'' \times 4.4''$.}
\label{fig:fig1}
\end{figure}

\begin{figure*}
\includegraphics[scale=0.28]{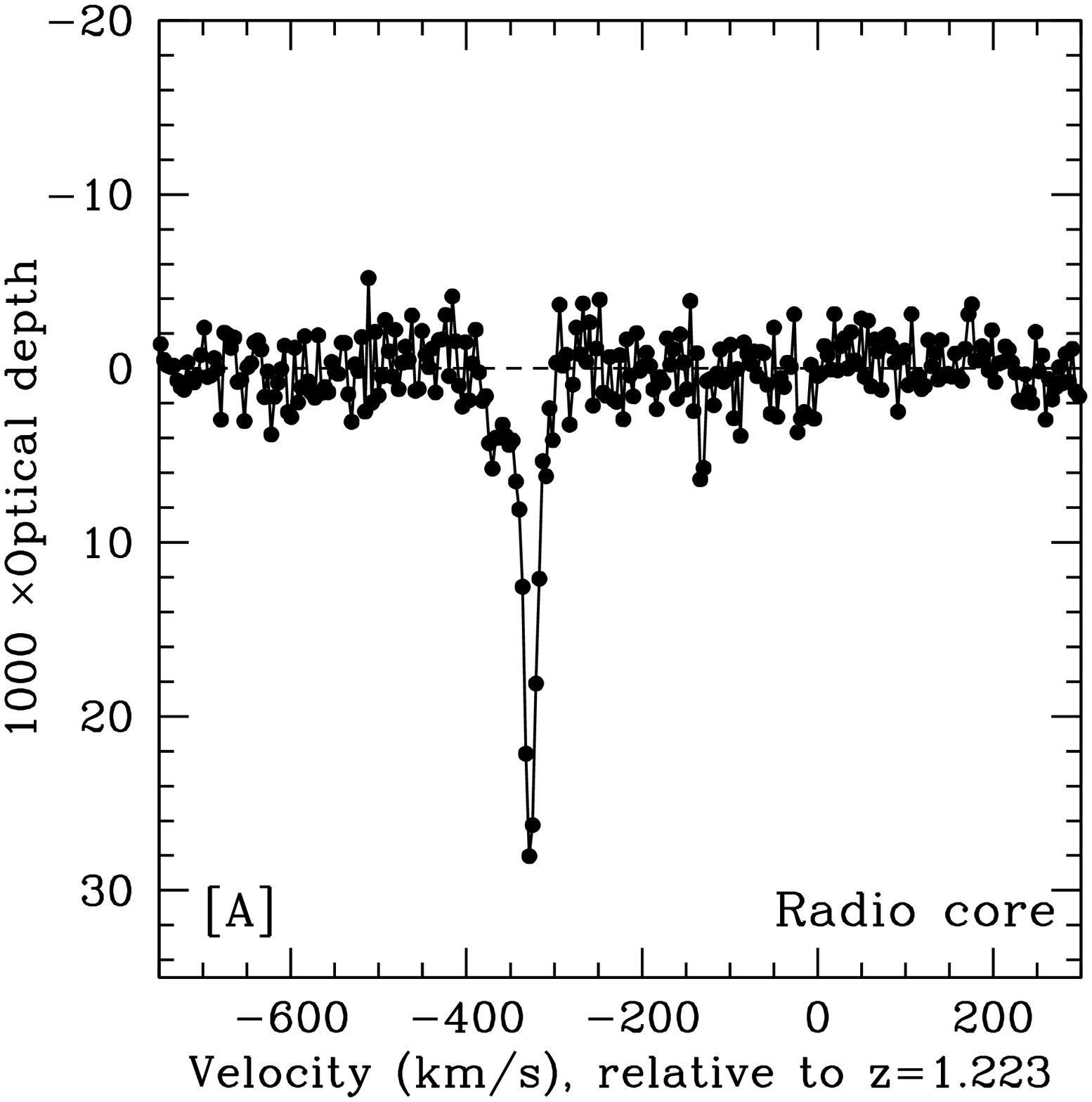}
\includegraphics[scale=0.28]{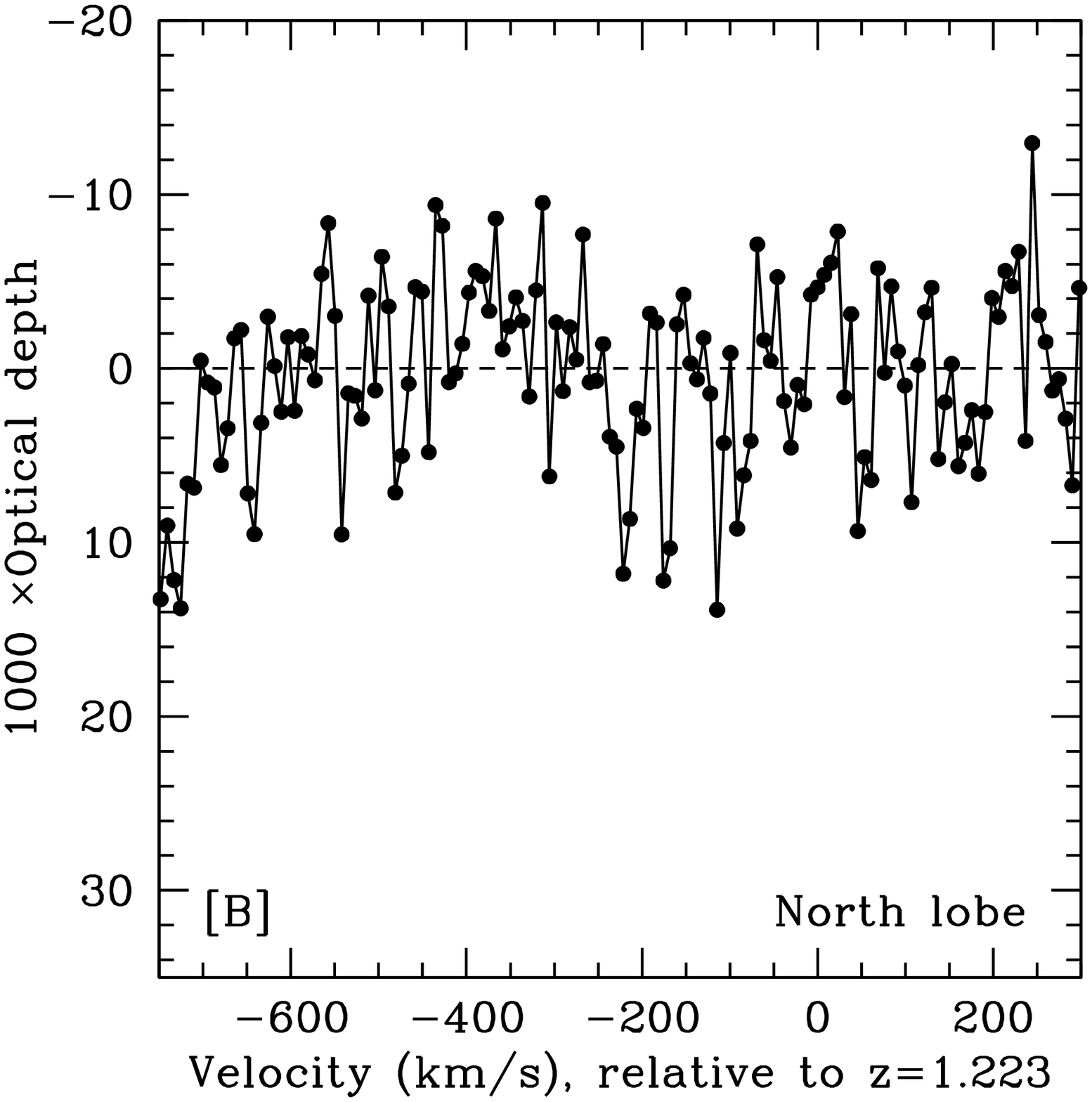}
\includegraphics[scale=0.28]{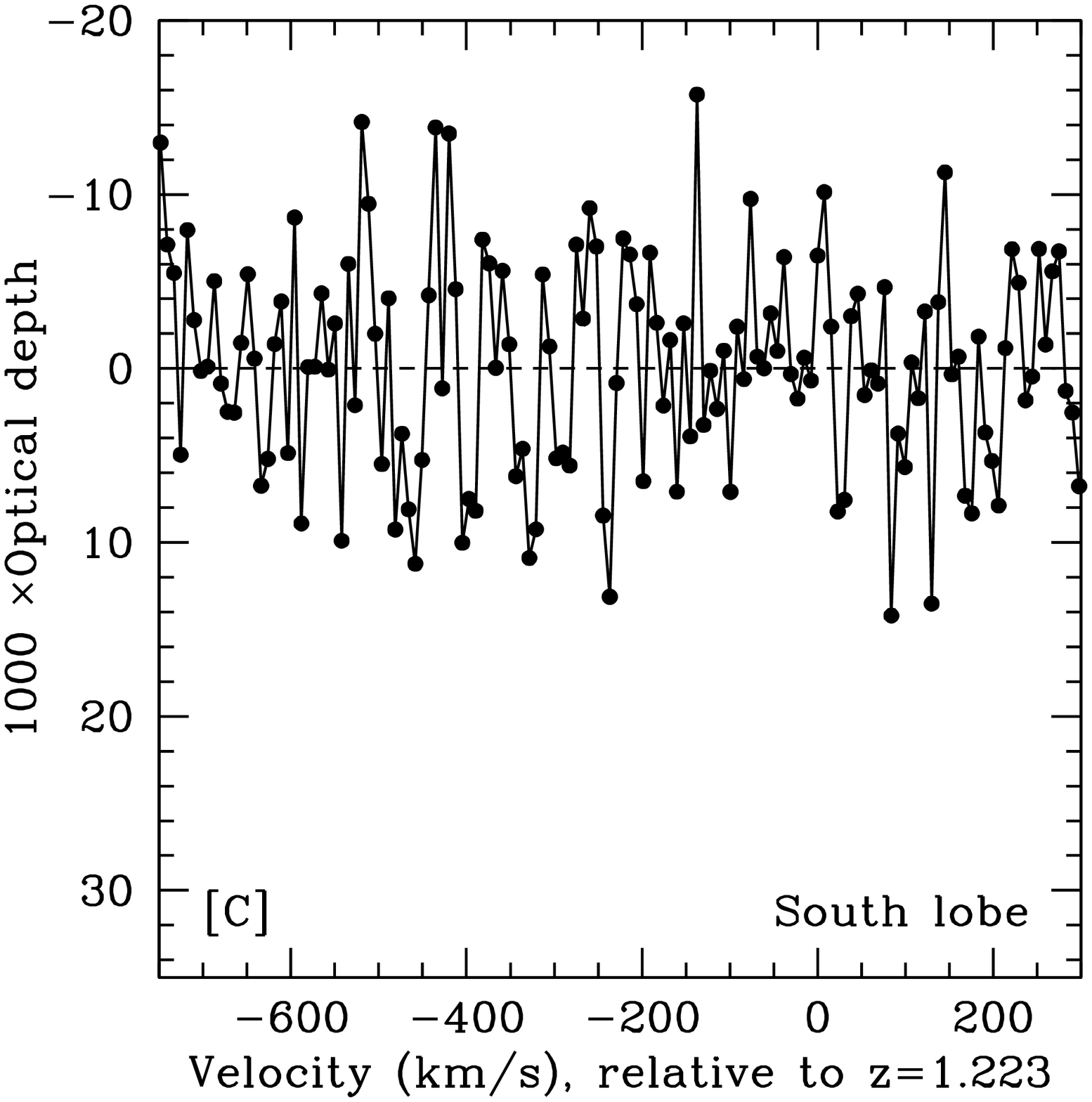}
\caption{GMRT \hii\ absorption spectra towards [A]~(left panel)~the radio core, 
[B]~(middle panel)~the northern lobe, and [C]~(right panel)~the southern lobe.  
In all three spectra, optical depth (in units of $1000 \times \tau$) is plotted 
against velocity, in km~s$^{-1}$, relative to the AGN redshift of $z = 1.2230$. 
It is clear that the optical depth sensitivity towards the two lobes is sufficient 
to detect \hii\ absorption of the same strength as that seen towards the radio core.}
\label{fig:fig2}
\end{figure*}


\section{The target: TXS\,1954+513}

TXS\,1954+513 is classified as a blazar in the literature \citep[e.g.][]{xie07,healey08,massaro09}, based
on its optical, X-ray, and radio characteristics. Specifically, the optical spectrum shows broad emission lines 
\citep[][]{lawrence96}, the source has a high X-ray luminosity \citep[$> 10^{43}$~erg~s$^{-1}$; ][]{britzen07}, 
the low-frequency radio spectrum is very flat (with a spectral index of $\alpha = 0.01$ between 1.4~GHz and 
5~GHz), and the radio flux density varies on timescales of a few days \citep{heeschen84}, all typical 
blazar properties. 

The redshift of TXS\,1954+513 has been estimated to be $z = 1.2230$ \citep{lawrence96}, based on the 
detection of a slew of both broad and narrow emission lines (including O{\sc ii}$\lambda$3727, 
O{\sc iii}$\lambda$4363, H$\gamma$, H$\delta$, etc). The large number of detected lines allow an accurate 
estimate of the AGN redshift: \citet{lawrence96} estimate the redshift uncertainty to be $\approx 0.0001$ 
for this source. 

Unfortunately, there are few optical photometric studies of TXS\,1954+513 in the literature. \citet{monet03}
obtain R$=17.34$ and B$=18.87$ from the second epoch of the USNO-B catalogue, with slightly different values
(R$=17.47$ and B$=18.44$) in the first epoch. This would imply a high inferred rest-frame 1216~\AA\ luminosity
\citep[e.g.][]{curran08}. However, the errors on the above measurements are uncertain \citep{monet03}; we 
hence chose to also carry out optical imaging of TXS\,1954+513 to obtain accurate photometry.

\section{Observations, Data Analysis and Results}

\subsection{GMRT imaging and spectroscopy}

The GMRT was initially used to search for associated redshifted \hii\ absorption towards 
TXS\,1954+513 in 2014 April, using the 610~MHz receivers and the GMRT software correlator as the 
backend. A bandwidth of 16.7~MHz was used for the observations, centred at the expected redshifted 
\hii\ line frequency ($\approx 638.96$~MHz) and subdivided into 512 channels; this yielded a 
velocity resolution of $\approx 15.3$~\kms\ and a total velocity coverage of $\approx 7815$~\kms\
(covering redshifts $z \approx 1.194 - 1.252$). Observations of the standard calibrators 3C286 
and 3C295 were used to calibrate the flux density scale and the system passband. No phase 
calibrator was observed as TXS\,1954+513 is known to be relatively compact at low frequencies. 
The on-source time was $\approx 60$~minutes.

The spectrum from the original GMRT observations showed statistically significant 
evidence of a narrow absorption feature close to the expected redshifted \hii\ line frequency. 
We hence re-observed TXS\,1954+513 in 2015~May to confirm the detection, again using 
the GMRT software backend as the correlator, but this time with a bandwidth of 4.17~MHz 
sub-divided into 512 channels, and centred at the redshifted \hii\ line frequency. 
This yielded a finer velocity resolution of $\approx 3.8$~\kms, allowing us to resolve 
out any narrow spectral components in the absorption profile. 3C286 was again used to 
calibrate the flux density scale and the system bandpass. The on-source time was 
$\approx 90$~minutes.

\subsection{Nickel Telescope photometry}

The field surrounding TXS\,1954+513 was imaged through the $B$ and $R$ filters using the 
1m Nickel telescope at Lick Observatory on UT 2016 Sep 14.  Images of 600s and 1800s 
were obtained under clear but humid conditions. Images of the standard fields PG2213-006
and SA~111 \citep{landolt09} were taken directly after the science field for photometric 
calibration.  Additional calibration frames (bias, dome flats) were also obtained.
Image processing of the on-sky images proceeded in standard fashion (e.g.\ bias subtraction,
flat field normalization) with custom scripts.  

We measured the counts in $14''$ diameter apertures around each calibration star and 
estimated the sky counts from a nearby, off-source aperture.  The measured zero points
corresponding to one count per second are $ZP_B = 22.1$\,mag and $ZP_R = 22.7$\,mag,
for the $B$ and $R$ filters, respectively.  Aperture photometry was then performed on TXS\,1954+513;
after applying the zero-point corrections, we obtained $B = 18.0$\,mag and 
$R = 16.9$\,mag, on the Vega magnitude scale. We estimate an uncertainty of 0.15\,mag 
from calibration errors and photon statistics.

%


\subsection{Data analysis and Results}

The GMRT data on TXS\,1954+513 were analysed in ``classic'' {\sc aips} following standard procedures. The data
were first carefully inspected, and edited to remove non-working antennas and time-specific intermittent bad
data (e.g. due to radio frequency interference, RFI). The antenna bandpasses were estimated from the data on
the bright calibrators 3C286 and 3C295. An initial estimate of the complex antenna gains was obtained by assuming 
TXS\,1954+513 to be a point source, and the gains were then refined via an iterative self-calibration procedure 
consisting of 4 rounds of phase-only self-calibration and imaging, followed by 2 rounds of amplitude-and-phase 
self-calibration and imaging. The amplitude-and-phase self-calibration step was followed by further data editing, 
based on the residuals after subtracting the continuum image from the calibrated visibilities. This iterative 
procedure was carried out until it yielded an image that did not improve upon further self-calibration and 
data editing. The final continuum image was then subtracted out from the calibrated spectral-line visibilities;
any remaining continuum emission was removed by fitting a first-order polynomial to line-free channels in 
each visibility spectrum, and subtracting this out. The residual visibilities were then shifted to the heliocentric
frame, and imaged to obtain the final spectral cube.  

The final GMRT 640~MHz continuum image of TXS\,1954+513, displayed in Fig.~\ref{fig:fig1}, shows a central core and 
two emission components extended in the north and south directions. A 3-component Gaussian model was fitted to the 
GMRT image (using the task {\sc jmfit}) to measure the flux densities of the three components; this yielded flux 
densities of $(1254.1 \pm 1.2)$~mJy (core), $(403.4 \pm 1.6)$~mJy (northern lobe) and $(198.1 \pm 1.3)$~mJy 
(southern lobe). The core emission is marginally resolved, with a peak flux of $(1221.30 \pm 0.69)$~mJy/Beam,
the northern lobe is highly resolved, with a peak flux of $(247.90 \pm 0.66)$~mJy/Bm, while the southern lobe
is also marginally resolved, with a peak flux of $(184.00 \pm 0.69)$~mJy/Bm. 

The radio core dominates the emission, providing $\approx 70$\% of the total flux density at our 
observing frequency of 640~MHz; as a result, the spectrum remains relatively flat at low frequencies, with a 
spectral index $\alpha \approx -0.2$ between 640~MHz and 1.4~GHz \citep{condon98} obtained from the total 
flux densities at the two frequencies (assuming the flux density $S_\nu$ at a frequency $\nu$ to be given 
by $S_\nu \propto \nu^\alpha$). The angular size of the source in the north-south direction is 
$\approx 20''$, i.e. $\approx 170$~kpc at the AGN redshift of $z = 1.2230$.

The three panels of Fig.~\ref{fig:fig2} show \hii\ spectra at the locations of the peaks of the three 
emission components of TXS\,1954+513 (after subtracting out a 2nd-order or 3rd-order baseline fitted to line-free
regions), with optical depth (in units of $1000 \times \tau$) plotted versus velocity (in \kms), relative to 
the heliocentric redshift $z = 1.2230$. These were obtained from the second observing run in 2015~May, whose data 
have both a better spectral 
resolution and a higher sensitivity. Strong \hii\ absorption is clearly visible in the spectrum through the radio 
core, shown in the left panel, while no \hii\ absorption was detected in the spectra through the two lobes, shown 
in the middle and right panels of the figure. Note that the optical depth sensitivities of the spectra towards the 
peaks of the northern and southern lobes are sufficient to detect \hii\ absorption of the same strength as that of 
the main absorption component seen against the core, but not to detect the weaker absorption component. We 
can thus rule out the possibility that the ``cloud'' producing the strong \hii\ absorption covers both the core 
and either of the lobes, but cannot rule out the possibility that the gas producing the weaker \hii\ absorption 
might cover all three source components.

The root-mean-square noise on the spectrum through the core was measured 
(from line-free channels) to be $\approx 2.3$~mJy per 3.8~\kms\ channel. The integrated \hii\ optical depth 
of the absorption line detected against the core is $(0.716 \pm 0.037)$~\kms, which implies an \hi\ column density of 
N$_{\rm HI} = (1.305 \pm 0.067) \times ({\rm T_s}/{\rm 100\:K}) \times 10^{20}$~cm$^{-2}$, for an assumed spin 
temperature of 100~K \citep[this value was used for consistency with the literature; e.g. ][]{vermeulen03,gupta06}.
Note that the spin temperature of neutral hydrogen in AGN environments could be much higher than 
the assumed 100~K \citep[e.g.][]{maloney96}; the above estimate of the \hi\ column density is hence likely 
to be a lower limit.

The detected \hii\ absorption towards the core of TXS\,1954+513 is relatively narrow, with a velocity width 
between 20\% points of $\approx 40$~\kms. No evidence is seen for a wide absorption component. The peak of 
the absorption is blueshifted from the AGN redshift \citep[$z= 1.2230 \pm 0.0001$; ][]{lawrence96} by 
$\approx 328$~\kms.

\section{Discussion}

A clear core-jet structure is seen on scales of tens of milli-arcseconds in the 1.6~GHz VLBI image of 
TXS\,1954+513, with a single jet in the eastward direction \citep[][]{polatidis95}.  The one-sided nature of 
the jet indicates that it is likely to point close to our line of sight to the core. Interestingly, however, our
GMRT 640~MHz continuum image shows that the radio structure is very different on arcsecond scales, with a core 
and two weaker lobes symmetrically extended in the north-south direction with a transverse angular 
extent of $\approx 25''$; this three-component north-south structure is also seen in the 1.4 GHz Very Large Array 
image of \citet{xu95}. The arcsecond-scale structure is thus of the ``core-dominated triple'' type \citep[e.g.][]{marecki06}.  
However, the milli-arcsecond scale core-jet structure is nearly perpendicular to the north-south arcsecond-scale 
structure. Such strong misalignments between the milli-arcsecond-scale and the arcsecond-scale radio structure have 
been observed earlier in a number of AGNs, especially BL~Lac objects \citep[e.g.][]{pearson88,wehrle92,conway93}, 
and, more recently, in a few core-dominated triples \citep[e.g.][]{marecki06}. Models that can account for these 
misalignments include twisted radio jets, either due to interaction with the ambient medium or precession of the 
axis of the central super-massive black hole \citep[e.g.][]{conway93,appl96}, and restarted AGN activity accompanied 
by a flip in the spin of the central black hole, possibly due to a galaxy merger event \citep[e.g.][]{marecki06}. 

We followed the prescription of \citet{curran08} to estimate the rest-frame UV luminosity of TXS\,1954+513, 
using a power-law model to extrapolate from its measured R-band and B-band magnitudes to infer its flux density 
at $1216 \times (1+z)$~\AA\ ($z = 1.223$), and, thence, its luminosity at this wavelength. Our measured 
R-band and B-band magnitudes are $R=16.9 \pm 0.15$ and $B=18.0 \pm 0.15$. Extrapolating from these measurements,
we obtain $L_{\rm UV} \approx (4.1 \pm 1.2) \times 10^{23}$~W~Hz$^{-1}$; this is by far the highest UV luminosity 
at which associated \hii\ absorption has been detected at any redshift \citep[e.g.][]{curran08,curran10}. This 
is also the first case of a detection of \hii\ absorption above the UV luminosity threshold suggested by 
\citet{curran10} of $L_{\rm UV} = 10^{23}$~W~Hz$^{-1}$. \citet{curran12b} use a simple model to argue that this 
is the UV luminosity threshold above which a quasar would ionize all the neutral gas in its host galaxy, for a 
range of gas densities. Our detection of \hii\ absorption shows 
that neutral hydrogen can indeed survive in AGN environments at high UV luminosities, $\gg 10^{23}$~W~Hz$^{-1}$. 

The peak of the \hii\ absorption is blueshifted by $\approx 328$~\kms\ from the AGN redshift of $1.2230$,
indicating that this is a case of gas outflow from an AGN. The detected absorption against the core of the 
GMRT 640~MHz image may arise against either the VLBI core or the VLBI-scale, eastward-pointing radio jet 
of TXS\,1954+513. However, since the jet direction is likely to lie close to our line of sight to the core, 
in both cases the gas cloud appears to be moving in the direction of the VLBI radio jet and away from the 
core. It is possible that the neutral gas in the vicinity of the AGN is being driven outward by the ram 
pressure of the radio jet, as has been suggested by recent numerical simulations \citep[e.g.][]{wagner11,wagner12}.

In summary, we have used the GMRT 610~MHz receivers to obtain only the fifth detection of associated redshifted 
\hii\ absorption at high redshifts, $z > 1$, in the blazar TXS\,1954+513. The AGN has a core-dominated triple structure 
in the GMRT image, with the two radio lobes oriented nearly perpendicular to the VLBI-scale radio jet; this
misalignment may arise due to either a twisted radio jet or restarted AGN activity with a spin-flip of the 
central black hole. The \hii\ absorption is detected towards the radio core of the GMRT 640~MHz image, i.e. 
towards either the VLBI core or the VLBI-scale radio jet, and is blueshifted by $\approx 328$~\kms\ from 
the AGN redshift, suggesting that the \hi\ cloud causing the absorption is being driven outward by the 
ram pressure of the jet. We have also used Nickel Telescope R- and B-band photometry of TXS\,1954+513 
to infer a high rest-frame 1216~\AA\ luminosity of $(4.1 \pm 1.2) \times 10^{23}$~W~Hz$^{-1}$ for the blazar. 
This is the first case of associated \hii\ absorption detected from an AGN with a rest-frame UV 
luminosity $> 10^{23}$~W~Hz$^{-1}$, showing that neutral gas can survive in the vicinity of active galactic 
nuclei with high ultraviolet luminosities.  

\section*{Acknowledgements}
We thank the staff of the GMRT who have made these observations possible. The GMRT is run by the National Centre 
for Radio Astrophysics of the Tata Institute of Fundamental Research. NK acknowledges support 
from the Department of Science and Technology via a Swarnajayanti Fellowship (DST/SJF/PSA-01/2012-13).


\bibliographystyle{mnras}
\bibliography{ms}

\begin{thebibliography}{}
\makeatletter
\relax
\def\mn@urlcharsother{\let\do\@makeother \do\$\do\&\do\#\do\^\do\_\do\%\do\~}
\def\mn@doi{\begingroup\mn@urlcharsother \@ifnextchar [ {\mn@doi@}
  {\mn@doi@[]}}
\def\mn@doi@[#1]#2{\def\@tempa{#1}\ifx\@tempa\@empty \href
  {http://dx.doi.org/#2} {doi:#2}\else \href {http://dx.doi.org/#2} {#1}\fi
  \endgroup}
\def\mn@eprint#1#2{\mn@eprint@#1:#2::\@nil}
\def\mn@eprint@arXiv#1{\href {http://arxiv.org/abs/#1} {{\tt arXiv:#1}}}
\def\mn@eprint@dblp#1{\href {http://dblp.uni-trier.de/rec/bibtex/#1.xml}
  {dblp:#1}}
\def\mn@eprint@#1:#2:#3:#4\@nil{\def\@tempa {#1}\def\@tempb {#2}\def\@tempc
  {#3}\ifx \@tempc \@empty \let \@tempc \@tempb \let \@tempb \@tempa \fi \ifx
  \@tempb \@empty \def\@tempb {arXiv}\fi \@ifundefined
  {mn@eprint@\@tempb}{\@tempb:\@tempc}{\expandafter \expandafter \csname
  mn@eprint@\@tempb\endcsname \expandafter{\@tempc}}}

\bibitem[\protect\citeauthoryear{{Aditya}, {Kanekar}  \& {Kurapati}}{{Aditya}
  et~al.}{2016}]{aditya16}
{Aditya} J.~N.~H.~S.,  {Kanekar} N.,   {Kurapati} S.,  2016, MNRAS, 455, 4000

\bibitem[\protect\citeauthoryear{{Appl}, {Sol}  \& {Vicente}}{{Appl}
  et~al.}{1996}]{appl96}
{Appl} S.,  {Sol} H.,   {Vicente} L.,  1996, A\&A, 310, 419

\bibitem[\protect\citeauthoryear{Barthel}{Barthel}{1989}]{barthel89}
Barthel P.~D.,  1989, ApJ, 336, 606

\bibitem[\protect\citeauthoryear{{Britzen} et~al.,}{{Britzen}
  et~al.}{2007}]{britzen07}
{Britzen} S.,  et~al., 2007, A\&A, 472, 763

\bibitem[\protect\citeauthoryear{{Carilli} {et al.}}{{Carilli} {et
  al.}}{2007}]{carilli07}
{Carilli} C.~L.,  {et al.} 2007, ApJ, 666, L9

\bibitem[\protect\citeauthoryear{{Condon}, {Cotton}, {Greisen}, {Yin},
  {Perley}, {Taylor}  \& {Broderick}}{{Condon} et~al.}{1998}]{condon98}
{Condon} J.~J.,  {Cotton} W.~D.,  {Greisen} E.~W.,  {Yin} Q.~F.,  {Perley}
  R.~A.,  {Taylor} G.~B.,   {Broderick} J.~J.,  1998, AJ, 115, 1693

\bibitem[\protect\citeauthoryear{{Conway} \& {Blanco}}{{Conway} \&
  {Blanco}}{1995}]{conway95}
{Conway} J.~E.,  {Blanco} P.~R.,  1995, ApJ, 449, L131

\bibitem[\protect\citeauthoryear{{Conway} \& {Murphy}}{{Conway} \&
  {Murphy}}{1993}]{conway93}
{Conway} J.~E.,  {Murphy} D.~W.,  1993, ApJ, 411, 89

\bibitem[\protect\citeauthoryear{{Curran} \& {Whiting}}{{Curran} \&
  {Whiting}}{2012}]{curran12b}
{Curran} S.~J.,  {Whiting} M.~T.,  2012, ApJ, 759, 117

\bibitem[\protect\citeauthoryear{{Curran}, {Whiting}, {Wiklind}, {Webb},
  {Murphy}  \& {Purcell}}{{Curran} et~al.}{2008}]{curran08}
{Curran} S.~J.,  {Whiting} M.~T.,  {Wiklind} T.,  {Webb} J.~K.,  {Murphy}
  M.~T.,   {Purcell} C.~R.,  2008, MNRAS, 391, 765

\bibitem[\protect\citeauthoryear{{Curran}, {Tzanavaris}, {Darling}, {Whiting},
  {Webb}, {Bignell}, {Athreya}  \& {Murphy}}{{Curran} et~al.}{2010}]{curran10}
{Curran} S.~J.,  {Tzanavaris} P.,  {Darling} J.~K.,  {Whiting} M.~T.,  {Webb}
  J.~K.,  {Bignell} C.,  {Athreya} R.,   {Murphy} M.~T.,  2010, MNRAS, 402, 35

\bibitem[\protect\citeauthoryear{{Curran}, {Whiting}, {Sadler}  \&
  {Bignell}}{{Curran} et~al.}{2013}]{curran13}
{Curran} S.~J.,  {Whiting} M.~T.,  {Sadler} E.~M.,   {Bignell} C.,  2013,
  MNRAS, 428, 2053

\bibitem[\protect\citeauthoryear{{Ger{\'e}b}, {Maccagni}, {Morganti}  \&
  {Oosterloo}}{{Ger{\'e}b} et~al.}{2015}]{gereb15}
{Ger{\'e}b} K.,  {Maccagni} F.~M.,  {Morganti} R.,   {Oosterloo} T.~A.,  2015,
  A\&A, 575, 44

\bibitem[\protect\citeauthoryear{{Gupta}, {Salter}, {Saikia}, {Ghosh}  \&
  {Jeyakumar}}{{Gupta} et~al.}{2006}]{gupta06}
{Gupta} N.,  {Salter} C.~J.,  {Saikia} D.~J.,  {Ghosh} T.,   {Jeyakumar} S.,
  2006, MNRAS, 373, 972

\bibitem[\protect\citeauthoryear{{Healey} et~al.,}{{Healey}
  et~al.}{2008}]{healey08}
{Healey} S.~E.,  et~al., 2008, ApJS, 175, 97

\bibitem[\protect\citeauthoryear{{Heeschen}}{{Heeschen}}{1984}]{heeschen84}
{Heeschen} D.~S.,  1984, AJ, 89, 1111

\bibitem[\protect\citeauthoryear{Ishwara-Chandra, Dwarakanath  \&
  Anantharamaiah}{Ishwara-Chandra et~al.}{2003}]{ishwar03}
Ishwara-Chandra C.~H.,  Dwarakanath K.~S.,   Anantharamaiah K.~R.,  2003,
  JA\&A, 24, 37

\bibitem[\protect\citeauthoryear{{Landolt}}{{Landolt}}{2009}]{landolt09}
{Landolt} A.~U.,  2009, AJ, 137, 4186

\bibitem[\protect\citeauthoryear{{Lawrence}, {Zucker}, {Readhead}, {Unwin},
  {Pearson}  \& {Xu}}{{Lawrence} et~al.}{1996}]{lawrence96}
{Lawrence} C.~R.,  {Zucker} J.~R.,  {Readhead} A.~C.~S.,  {Unwin} S.~C.,
  {Pearson} T.~J.,   {Xu} W.,  1996, ApJS, 107, 541

\bibitem[\protect\citeauthoryear{{Maloney}, {Hollenbach}  \&
  {Tielens}}{{Maloney} et~al.}{1996}]{maloney96}
{Maloney} P.~R.,  {Hollenbach} D.~J.,   {Tielens} A.~G.~G.~M.,  1996, ApJ, 466,
  561

\bibitem[\protect\citeauthoryear{{Marecki}, {Thomasson}, {Mack}  \&
  {Kunert-Bajraszewska}}{{Marecki} et~al.}{2006}]{marecki06}
{Marecki} A.,  {Thomasson} P.,  {Mack} K.-H.,   {Kunert-Bajraszewska} M.,
  2006, A\&A, 448, 479

\bibitem[\protect\citeauthoryear{{Massaro}, {Giommi}, {Leto}, {Marchegiani},
  {Maselli}, {Perri}, {Piranomonte}  \& {Sclavi}}{{Massaro}
  et~al.}{2009}]{massaro09}
{Massaro} E.,  {Giommi} P.,  {Leto} C.,  {Marchegiani} P.,  {Maselli} A.,
  {Perri} M.,  {Piranomonte} S.,   {Sclavi} S.,  2009, A\&A, 495, 691

\bibitem[\protect\citeauthoryear{{Monet} {et al.}}{{Monet} {et
  al.}}{2003}]{monet03}
{Monet} D.~G.,  {et al.} 2003, AJ, 125, 984

\bibitem[\protect\citeauthoryear{{Moore}, {Carilli}  \& {Menten}}{{Moore}
  et~al.}{1999}]{moore99}
{Moore} C.~B.,  {Carilli} C.~L.,   {Menten} K.~M.,  1999, ApJ, 510, L87

\bibitem[\protect\citeauthoryear{{Morganti}, {Oosterloo}, {Tadhunter}, {van
  Moorsel}  \& {Emonts}}{{Morganti} et~al.}{2005a}]{morganti05b}
{Morganti} R.,  {Oosterloo} T.~A.,  {Tadhunter} C.~N.,  {van Moorsel} G.,
  {Emonts} B.,  2005a, A\&A, 439, 521

\bibitem[\protect\citeauthoryear{{Morganti}, {Tadhunter}  \&
  {Oosterloo}}{{Morganti} et~al.}{2005b}]{morganti05}
{Morganti} R.,  {Tadhunter} C.~N.,   {Oosterloo} T.~A.,  2005b, A\&A, 444, L9

\bibitem[\protect\citeauthoryear{{Morganti}, {Peck}, {Oosterloo}, {van
  Moorsel}, {Capetti}, {Fanti}, {Parma}  \& {de Ruiter}}{{Morganti}
  et~al.}{2009}]{morganti09}
{Morganti} R.,  {Peck} A.~B.,  {Oosterloo} T.~A.,  {van Moorsel} G.,  {Capetti}
  A.,  {Fanti} R.,  {Parma} P.,   {de Ruiter} H.~R.,  2009, A\&A, 505, 559

\bibitem[\protect\citeauthoryear{{Morganti}, {Fogasy}, {Paragi}, {Oosterloo}
  \& {Orienti}}{{Morganti} et~al.}{2013}]{morganti13}
{Morganti} R.,  {Fogasy} J.,  {Paragi} Z.,  {Oosterloo} T.,   {Orienti} M.,
  2013, Science, 341, 1082

\bibitem[\protect\citeauthoryear{{Pearson} \& {Readhead}}{{Pearson} \&
  {Readhead}}{1988}]{pearson88}
{Pearson} T.~J.,  {Readhead} A.~C.~S.,  1988, ApJ, 328, 114

\bibitem[\protect\citeauthoryear{{Peck} \& {Taylor}}{{Peck} \&
  {Taylor}}{2002}]{peck02}
{Peck} A.~B.,  {Taylor} G.~B.,  2002, New Astr. Rev., 46, 273

\bibitem[\protect\citeauthoryear{{Polatidis}, {Wilkinson}, {Xu}, {Readhead},
  {Pearson}, {Taylor}  \& {Vermeulen}}{{Polatidis} et~al.}{1995}]{polatidis95}
{Polatidis} A.~G.,  {Wilkinson} P.~N.,  {Xu} W.,  {Readhead} A.~C.~S.,
  {Pearson} T.~J.,  {Taylor} G.~B.,   {Vermeulen} R.~C.,  1995, ApJS, 98, 1

\bibitem[\protect\citeauthoryear{{Roberts}}{{Roberts}}{1970}]{roberts70}
{Roberts} M.~S.,  1970, ApJ, 161, L9

\bibitem[\protect\citeauthoryear{{Taylor}, {Vermeulen}, {Readhead}, {Pearson},
  {Henstock}  \& {Wilkinson}}{{Taylor} et~al.}{1996}]{taylor96}
{Taylor} G.~B.,  {Vermeulen} R.~C.,  {Readhead} A.~C.~S.,  {Pearson} T.~J.,
  {Henstock} D.~R.,   {Wilkinson} P.~N.,  1996, ApJS, 107, 37

\bibitem[\protect\citeauthoryear{Uson, Bagri  \& Cornwell}{Uson
  et~al.}{1991}]{uson91}
Uson J.~M.,  Bagri D.~S.,   Cornwell T.~J.,  1991, Phys.~Rev.~Lett., 67, 3328

\bibitem[\protect\citeauthoryear{{Vermeulen} et~al.,}{{Vermeulen}
  et~al.}{2003}]{vermeulen03}
{Vermeulen} R.~C.,  et~al., 2003, A\&A, 404, 861

\bibitem[\protect\citeauthoryear{{Wagner} \& {Bicknell}}{{Wagner} \&
  {Bicknell}}{2011}]{wagner11}
{Wagner} A.~Y.,  {Bicknell} G.~V.,  2011, ApJ, 728, 29

\bibitem[\protect\citeauthoryear{{Wagner}, {Bicknell}  \& {Umemura}}{{Wagner}
  et~al.}{2012}]{wagner12}
{Wagner} A.~Y.,  {Bicknell} G.~V.,   {Umemura} M.,  2012, ApJ, 757, 136

\bibitem[\protect\citeauthoryear{{Wehrle}, {Cohen}, {Unwin}, {Aller}, {Aller}
  \& {Nicolson}}{{Wehrle} et~al.}{1992}]{wehrle92}
{Wehrle} A.~E.,  {Cohen} M.~H.,  {Unwin} S.~C.,  {Aller} H.~D.,  {Aller} M.~F.,
    {Nicolson} G.,  1992, ApJ, 391, 589

\bibitem[\protect\citeauthoryear{{Xie}, {Dai}  \& {Zhou}}{{Xie}
  et~al.}{2007}]{xie07}
{Xie} G.~Z.,  {Dai} H.,   {Zhou} S.~B.,  2007, AJ, 134, 1464

\bibitem[\protect\citeauthoryear{{Xu}, {Readhead}, {Pearson}, {Polatidis}  \&
  {Wilkinson}}{{Xu} et~al.}{1995}]{xu95}
{Xu} W.,  {Readhead} A.~C.~S.,  {Pearson} T.~J.,  {Polatidis} A.~G.,
  {Wilkinson} P.~N.,  1995, ApJS, 99, 297

\bibitem[\protect\citeauthoryear{{van Gorkom}, {Knapp}, {Ekers}, {Ekers},
  {Laing}  \& {Polk}}{{van Gorkom} et~al.}{1989}]{vangorkom89}
{van Gorkom} J.~H.,  {Knapp} G.~R.,  {Ekers} R.~D.,  {Ekers} D.~D.,  {Laing}
  R.~A.,   {Polk} K.~S.,  1989, AJ, 97, 708

\makeatother
\end{thebibliography}

%

\bsp	
\label{lastpage}
\end{document}